\documentclass{article}

\usepackage[dvips]{graphicx}
\usepackage[dvips]{color}
\usepackage[thinspace,thinqspace]{fancyunits}
\usepackage{wrapfig}
\usepackage{booktabs}
\usepackage{nicefrac}
\usepackage{fancyhdr}   

\usepackage{lineno}
\usepackage{icrctc07}

\addunit{\eV}{eV}
\addunit{\EeV}{EeV}
\addunit{\yr}{yr}
\addunit{\VEM}{VEM}

\title{Measurement of the UHECR
  energy spectrum using data from the Surface Detector of the Pierre Auger
  Observatory} 
\shorttitle{Measurement of the energy spectrum using data from the Pierre Auger
  Observatory}
\authors{Markus Roth$^1$, for the
  Auger Collaboration$^2$}
\afiliations{
$^1$Institut f\"ur Kernphysik, Forschungszentrum Karlsruhe, POB
  3640, D-76021 Karlsruhe, Germany \\
$^2$Observatorio Pierre Auger, Av. San Mart\'in Norte 304, (5613) Malarg\"ue,
Mendoza, Argentina}
\email{Markus.Roth@ik.fzk.de}
\abstract{
At the southern site of the Pierre Auger Observatory,
which is close to completion, an exposure that significantly exceeds the
largest forerunner experiments has already been accumulated. We report a
measurement of the cosmic ray energy spectrum based on the high statistics
collected by the surface detector. The methods developed to determine the
spectrum from reconstructed observables are described. The energy calibration
of the observables, which exploits the correlation of surface detector data
with fluorescence measurements in hybrid events, is presented in
detail. 
The methods are simple and robust, exploiting the combination of fluorescence
detector (FD) and surface detector (SD) and do not rely on detailed
numerical simulation or any assumption about the chemical composition.
Besides presenting statistical uncertainties, we address the impact of
systematic uncertainties.  
}
\begin{document}
%
%
\maketitle
\section{Introduction}
The Pierre Auger Observatory~\cite{bib:AugerNIM04} is designed to measure the
extensive air showers produced by the highest energy cosmic rays ($E>
10^{18.5}$~eV) with the goal of discovering their origins and composition.
Two different techniques are used to detect air showers.  Firstly, a
collection of telescopes is used to sense the fluorescence light produced by
excitation of nitrogen induced by the cascade of particles in the atmosphere. 
The FD provides a nearly calorimetric, model-independent energy measurement,
because the fluorescence light is produced in proportion to energy dissipation
by a shower in the atmosphere~\cite{bib:RisseATP20,bib:Barbosa}. This
method can be used only when the sky is moonless and dark, and thus has
roughly a 10\% duty cycle~\cite{bib:ICRC07Dawson}. 
The second method uses an array of detectors on the ground to sample particle
densities as the air shower arrives at the Earth's surface. 
The surface detector has a 100\% duty cycle~\cite{bib:ICRC07Suomijarvi}. A
subsample of air showers detected by both instruments, dubbed hybrid events,
are very precisely measured~\cite{bib:ICRC07Perrone} and provide an invaluable
energy calibration tool.  
Hybrid events make it possible to relate the shower energy (FD) to the ground
parameter $S(1000)$. 
%
\section{Analysis procedure}
\begin{figure}[t]
    \centering
    \includegraphics[width=0.465\textwidth]{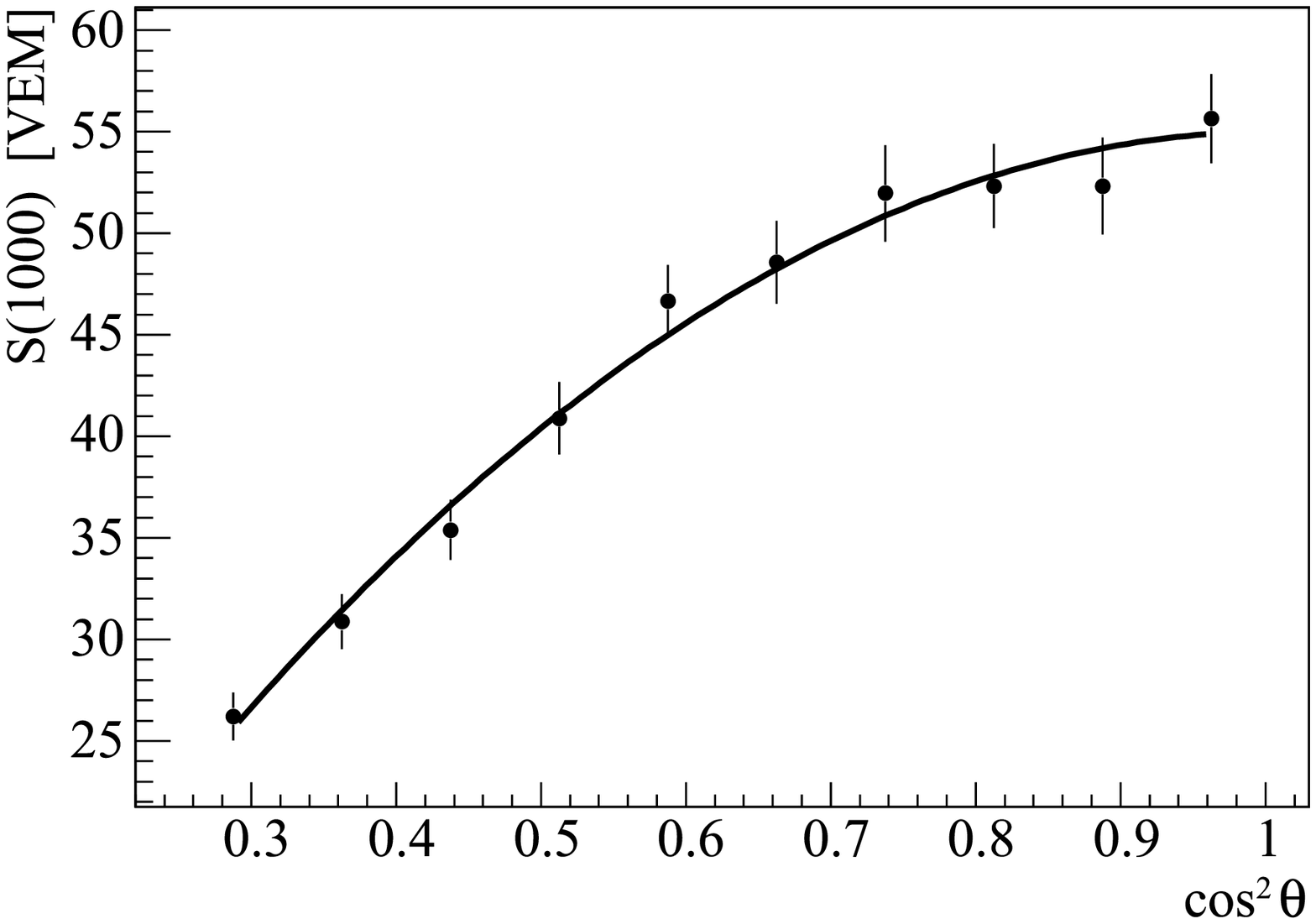}
    \vspace*{-0.3cm}
    \caption{Derived attenuation curve, $CIC(\theta)$, fitted with a quadratic
    function. 
      \label{fig:CorrectionCurve}} 
\end{figure}

%
%
%
The parameter $S(1000)$ characterises the energy of a cosmic ray shower
detected by the SD array and is the signal in units of \VEM{} that would
be produced in a tank at a distance of \unit{1000}{\meter} from the shower
axis. One \VEM{} is the signal produced by a single relativistic muon passing
vertically through the centre of a water tank. A likelihood method is applied
to obtain the lateral distribution function, where the shower axis, $S(1000)$
and the curvature of the shower front are
determined~\cite{bib:ICRC05Barnhill}. The 
selection criteria are such to ensure the rejection of accidental triggers 
(physics trigger) and the events are well contained in the SD array (quality
trigger), i.e.
we require that all six nearest neighbours of the station with the  
highest signal be active. In this way we guarantee that the core of the  
shower is contained inside the array and enough of the shower is  
sampled to make an S(1000) measurement.
\begin{figure}[t]
    \centering
    \includegraphics[width=0.47\textwidth]{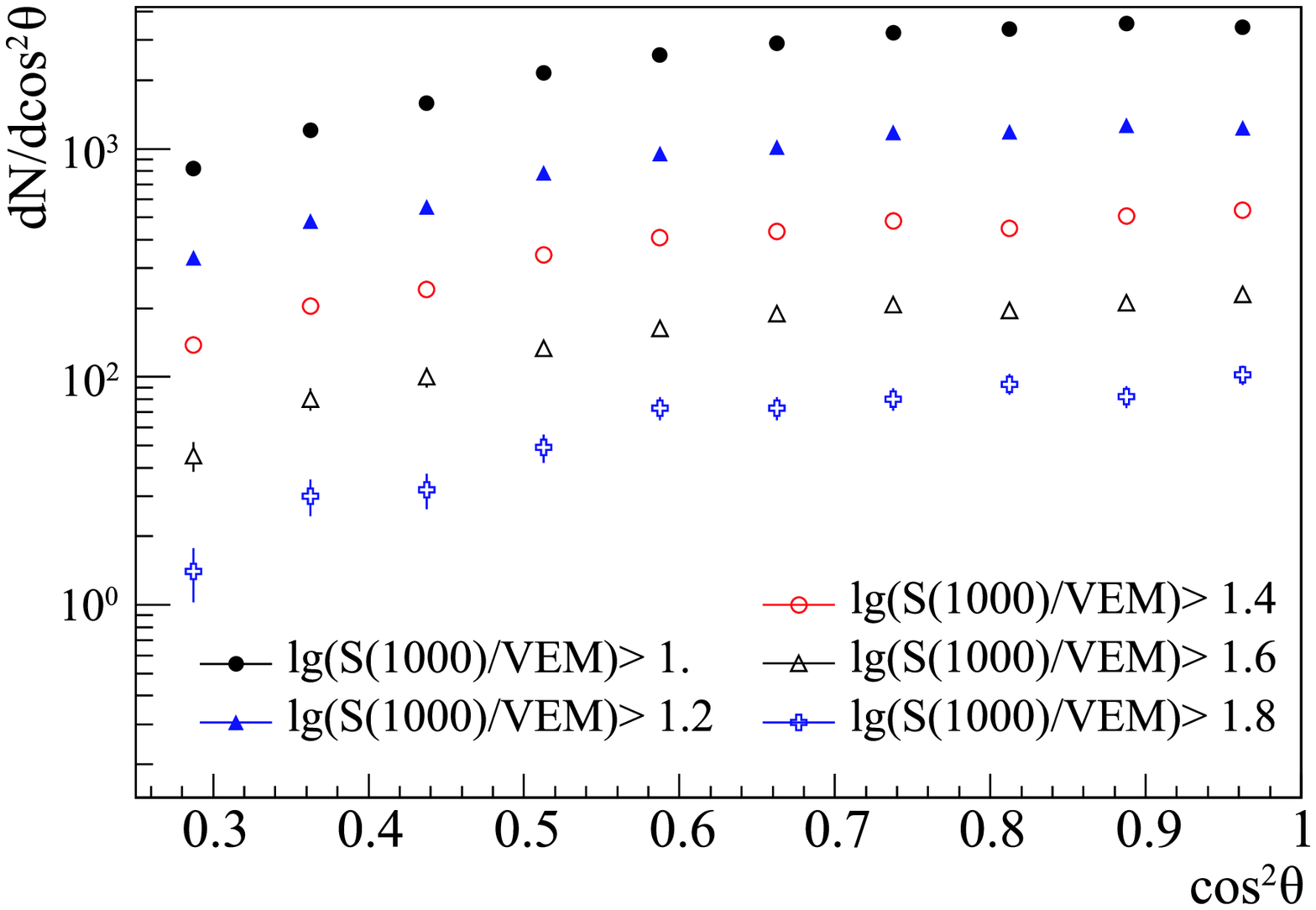}
    \vspace*{-0.3cm}
    \caption{Integral number of events vs $\cos^2\theta$ for the indicated
    minimum value of $S(1000)$.
      \label{fig:AttenuiationCurve}} 
\end{figure}
The present data set is taken from 1 January, 2004 through 28 February, 2007
while the array has been growing in size. To ensure an 
excellent data quality we remove periods with problems due to failures in data
acquisition, due to 
lightning and 
hardware difficulties. We select events only if the zenith angle is less than
$60\degree$ and the reconstructed energy is above $\unit{3}{\EeV}$. For this
analysis, the array
is fully efficient for detecting such showers, so the acceptance at any time
is solely determined by the geometric aperture of the
array~\cite{bib:ICRC05Allard}. The integrated 
exposure mounts up to about $\unit{5165}{\kilo\meter\Squared\;\sterad\;\yr}$,
which is a factor of more than 3 larger than the exposure obtained by the
largest forerunner experiment AGASA~\cite{bib:Takeda03}. Moreover the present
acceptance exceeds the one given
in~\cite{bib:ICRC05Sommers} by a factor of about 3.
For a given energy the value of $S(1000)$ decreases with zenith angle,
$\theta$, due to attenuation of the shower particles and geometrical
effects. Assuming an isotropic flux for the whole energy range considered,
i.e.~the intensity distribution is uniform when binned in $\cos^2\theta$, 
we extract the shape of the attenuation curve from the data.
In Figure~\ref{fig:AttenuiationCurve} several intensities, $I_i=I(>S_i(1000))$,
above a given value of $\lg S_i(1000)$ are shown as a function of
$\cos^2\theta$. The choice of the threshold $\lg S(1000)$ is not critical
since the shape is nearly the same
within the statistical limit. 
The fitted attenuation curve, $CIC(\theta)=1+a\;x+b\;x^2$, is a quadratic
function of $x=\cos^2\theta-\cos^238^\circ$ as displayed in
Figure~\ref{fig:CorrectionCurve} for a particular constant intensity cut,
$I_0=128 \text{ events}$, with $a=0.94\pm0.06$ and $b=-1.21\pm0.27$.  
The cut corresponds to a shower size of about $S_{38^\circ}=\unit{47}{\VEM}$ and
equivalently to an energy of about $\unit{9}{\EeV}$. Since the
average angle is $\langle \theta \rangle \simeq 38^\circ$ we take this angle
as reference and convert $S(1000)$ into $S_{38^\circ}$ by $S_{38^\circ}\equiv
S(1000)/CIC(\theta)$. It may be regarded as the signal $S(1000)$ the shower
would have produced had it arrived at $\theta=38^\circ$.
The reconstruction accuracy of the parameter $S(1000)$, $\sigma_{S(1000)}$,
comprises 3 contributions and these are taken into account in inferring
$S_{38^\circ}$ and its uncertainty $\sigma_{S_{38^\circ}}$: a statistical
uncertainty due to 
the finite size of the detector and the limited dynamic range of the signal
detection, a systematic uncertainty due to the assumptions of the shape of the
lateral distribution and finally due to the shower-to-shower
fluctuations~\cite{bib:ICRC07Ave}.   
To infer the energy we have to establish the relation between $S_{38^\circ}$
and the calorimetric energy measurement, $E_{FD}$. A 
set of hybrid events of high quality is selected based on the criteria
reported in~\cite{bib:ICRC07Perrone} without applying the cut on the field of
view, which appears to have a negligible effect for the topic addressed here. 
A small correction to account for the energy carried away by high energy muons
and neutrinos, the so-called \emph{invisible energy}, depends slightly on
mass and hadronic model. The applied correction is based on the average for
proton and iron showers simulated with the QGSJet model and sums up to about
$10\%$ and its systematic uncertainty contributes $4\%$ to the total
uncertainty in FD energy~\cite{bib:Barbosa}.
Moreover the SD quality cuts described above are applied.
The criteria include a measurement of the vertical aerosol optical depth 
profile (VAOD(h)) \cite{bib:ICRC07ben-zvi} using laser shots generated by the 
central laser facility (CLF) \cite{bib:JINST06Fick} and observed by the FD in
the same hour of each selected hybrid event. 
\begin{figure}[!t]
    \centering
    \includegraphics[width=0.46\textwidth]{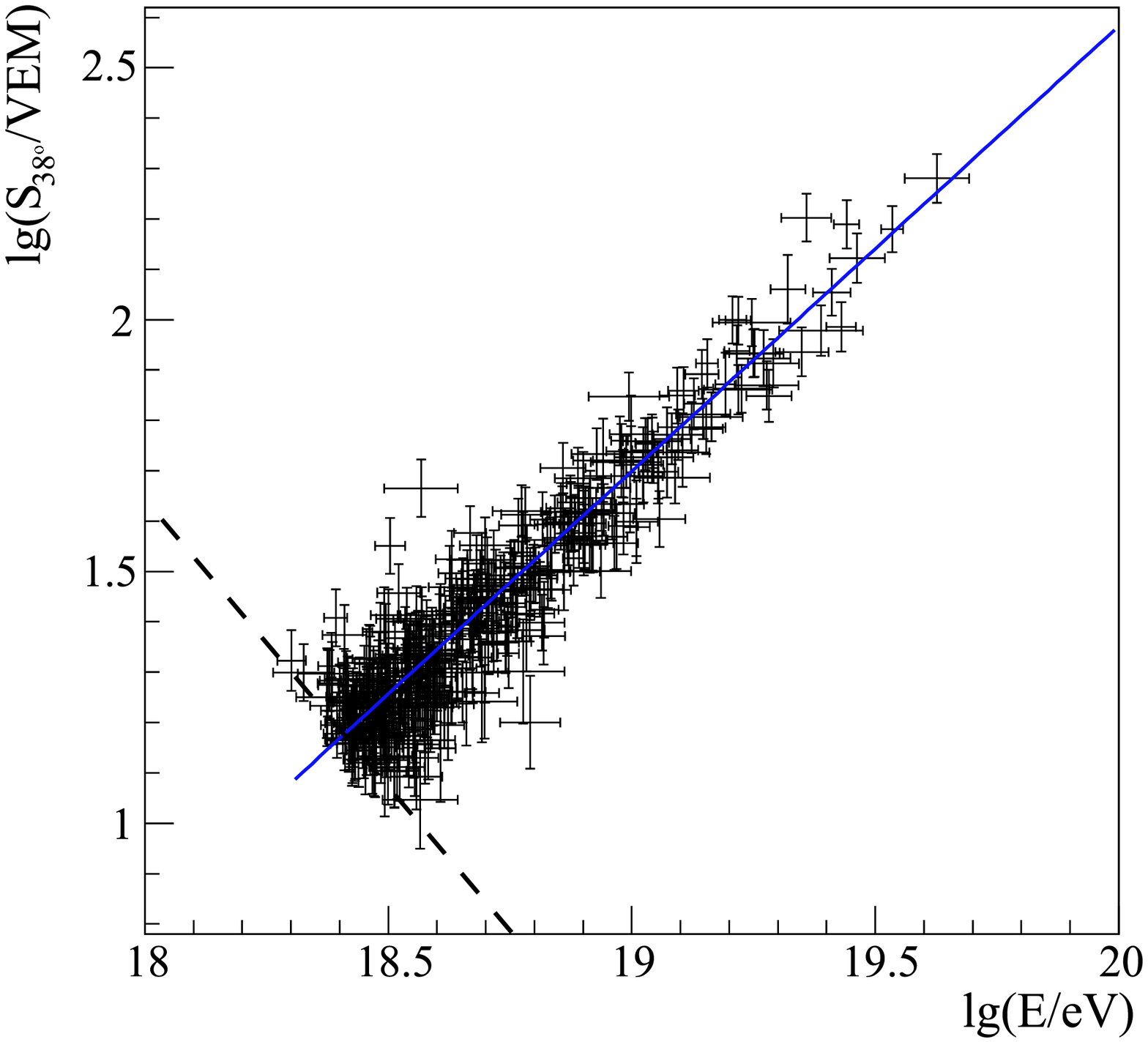}
    \vspace*{-0.3cm}
    \caption{Correlation between  $\lg E_{FD}$ and $\lg S_{38^{\circ}}$
      for the 387 hybrid events used in the fit. The full line is the best fit
      to the data. Events below the dashed line were not included in the fit.  
      \label{fig:SDCalibration}} 
\end{figure}
The selected hybrid events were used to calibrate the SD energy. The following
procedure was adopted. For each 
hybrid event, with measured FD energy $E_{FD}$, the SD energy estimator
$S_{38^\circ}$ was determined from the measured $S(1000)$ by using the
constant intensity method described above.   
For each event the uncertainty in $S_{38^{\circ}}$
is estimated by summing in quadrature three contributions:
the uncertainty in the 
constant intensity parametrization, $\sigma_{S_{38^{\circ}}}(CIC)$ , the
angular accuracy of the event,  $\sigma_{cos\theta}$, 
and the uncertainty in the measured $S(1000)$, $\sigma_{S(1000)}$.
The fluorescence yield used to estimate the energy $E_{FD}$  is taken
from~\cite{bib:naganoFY}. 
An uncertainty in the FD energy, $\sigma_{E_{FD}}$, was also assigned to each
event. Several sources were considered. The uncertainty in the hybrid shower
geometry, the statistical uncertainty in the Gaisser-Hillas fit to the profile
of the energy deposits and the
statistical uncertainty in the invisible energy correction were 
fully propageted.
The uncertainty in the VAOD measurement
was also propagated to the FD energy on an event-by-event basis, by evaluating
the FD energy shift obtained when changing the VAOD profile by its
uncertainty. These individual contributions were considered to be
uncorrelated, and were thus combined in quadrature to obtain
$\sigma_{E_{FD}}$.
The data appear to be well described by a linear relation 
$ \lg E_{FD} = A+ B \cdot \lg S_{38^{\circ}}$ (see Figure~\ref{fig:SDCalibration}).
A linear least square fit of the data was performed. To avoid
possible biases, low energy events, 
below the dashed line, which is orthogonal to the best fit
line and intersects it at $\lg (S_{38^{\circ}}=\unit{15}{\VEM})$, were not
included in 
the fit. 
\begin{figure}[!t]
    \centering
    \includegraphics[width=0.45\textwidth]{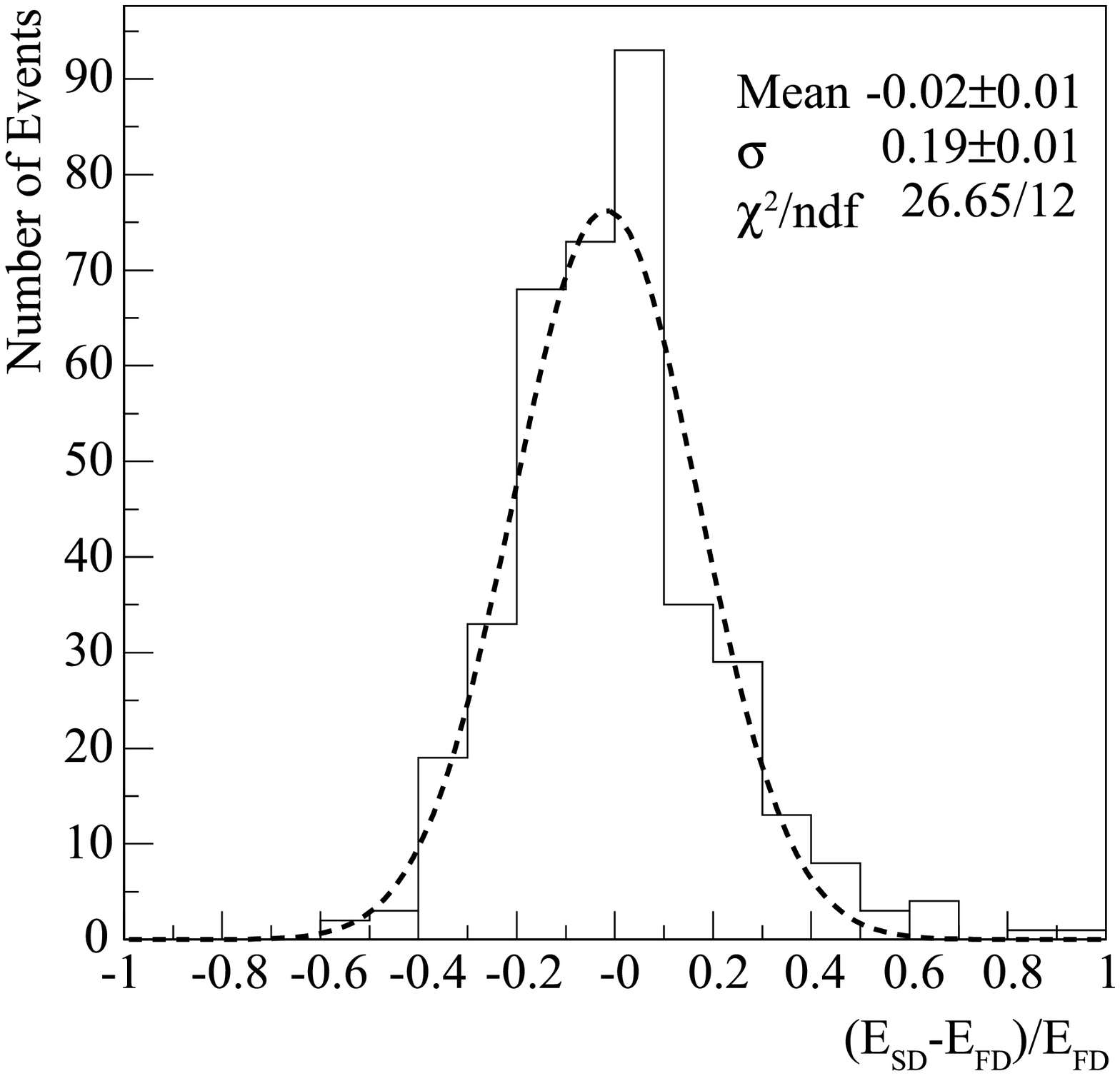}
    \vspace*{-0.3cm}
    \caption{Fractional difference between the FD and SD energy for the 387
      selected hybrid events.
      \label{fig:EnergyResolution}} 
\end{figure}

\begin{figure*}
  \centering
  \vspace*{-0.1cm}
  \begin {minipage}[b]{0.28\textwidth}
    \centering
    \caption{Auger spectrum $J$ as a function of energy. Vertical error bars
      represent the statistical uncertainty only. 
      The statistical  and systematic uncertainties in the energy
      scale are of the order of $\approx 6\%$ and $\approx 22\%$,
      respectively. 
      \label{fig:EnergySpectrum}} 
  \end{minipage}
  \begin {minipage}[b]{0.7\textwidth}
    \centering
    \includegraphics[width=1.\textwidth]{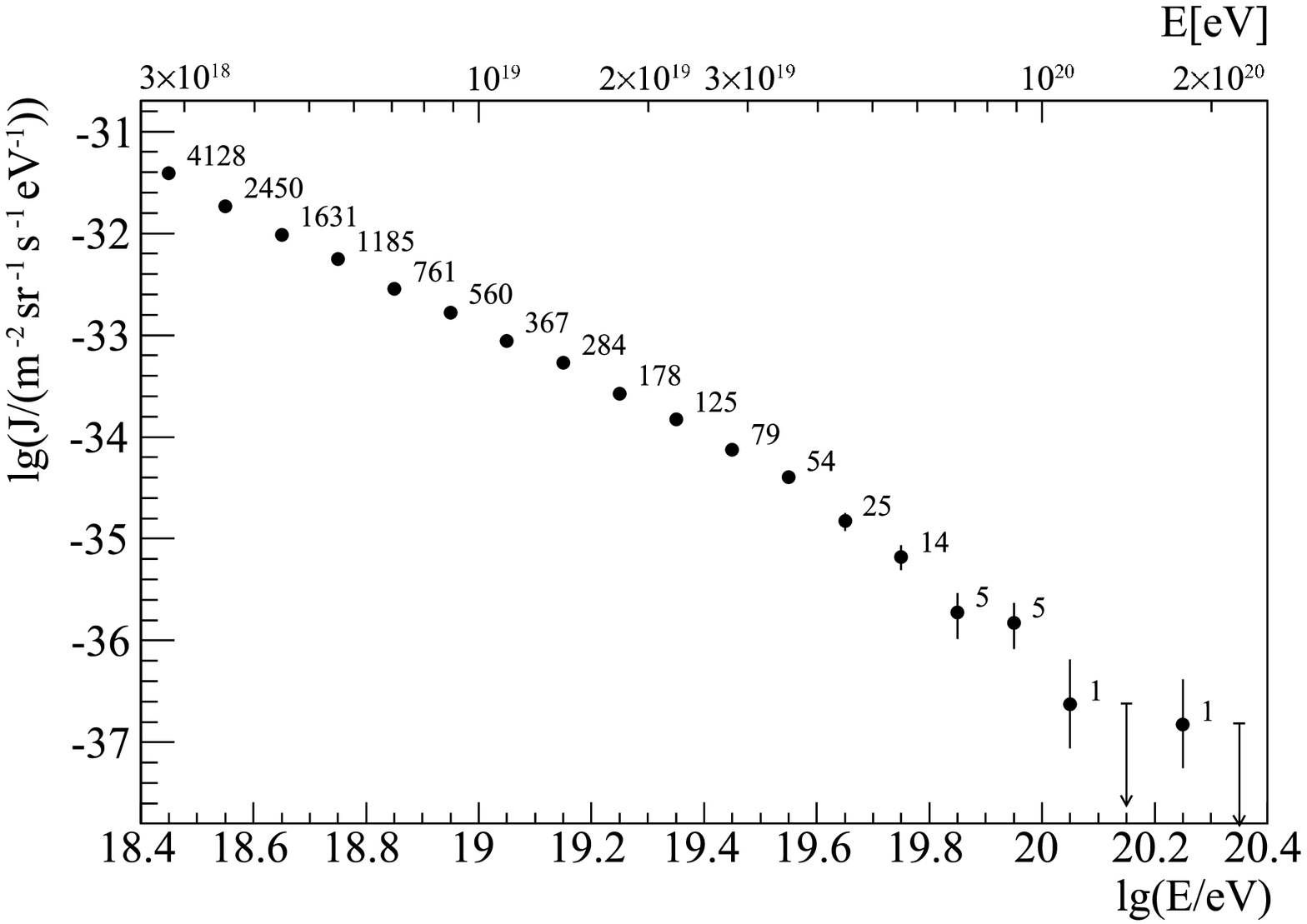}
  \end{minipage}
  \vspace*{-0.2cm}
\end{figure*}

\begin{figure}[h]
  \vspace*{-0.2cm}
    \centering
    \includegraphics[width=0.48\textwidth]{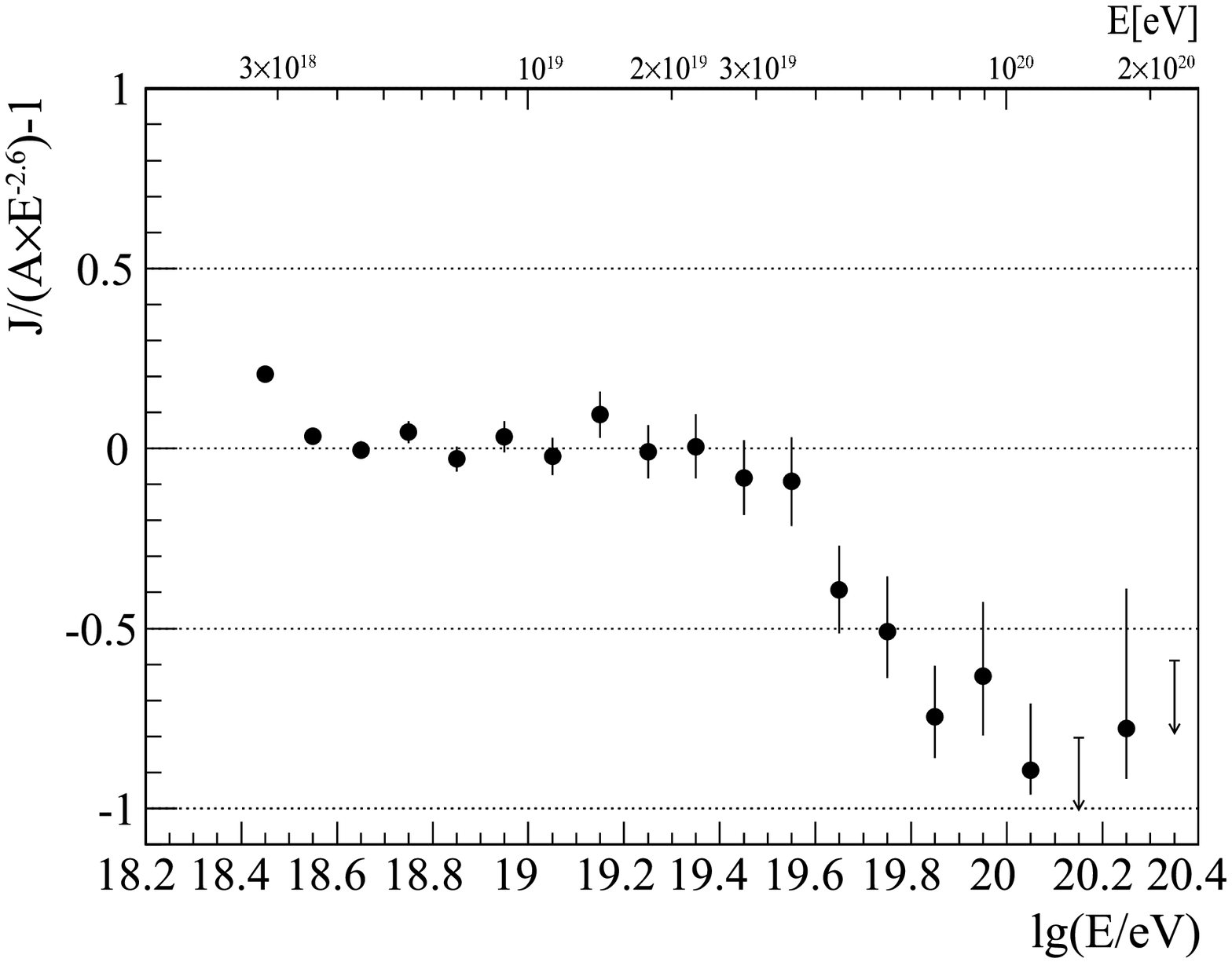}
    \vspace*{-0.3cm}
    \caption{Fractional difference between the derived spectrum and an assumed
    flux $\propto E^{-2.6}$ as a function of energy. 
      \label{fig:EnergyComparison}} 
\end{figure}

An iterative procedure was used to determine the dashed line, and it
was checked that the results of the fit were stable.  
The best fit yields $A=17.08\pm 0.03$ and $B=1.13\pm 0.02$ with a reduced
$\chi^2$ of 1.3 for
$
\lg E_{SD}=A + B\cdot \lg S_{38^\circ} \text{ in [eV]}.
$
The relative statistical uncertainty in the derived SD energy,
$\nicefrac{\sigma_{E_{SD}}}{ E_{SD}}$, is rather small, e.g.~of the order of
5\% at $\unit{\power{10}{20}}{\eV}$.
The energy spectrum $J$ is displayed in
Figure~\ref{fig:EnergySpectrum} together with its statistical uncertainty. The
individual systematic uncertainties in determining $E_{SD}$ coming from the FD
sum up to 22\%. For illustrative purposes we show in
Figure~\ref{fig:EnergyComparison} the difference of the flux with respect to an
assumed flux $\propto E^{-2.6}$. 
The largest uncertainties are given by the absolute
fluorescence yield (14\%), the absolute calibration of the FD (9.5\%) and the
reconstruction method (10\%). 
The uncertainty due to the dependence of the fluorescence spectrum on
pressure (1\%), humidity (5\%) and temperature (5\%) are take into account as
well as the wavelength dependent response of the FD, the aerosol phase
function, invisible energy and others, which are well below 4\%
(see~\cite{bib:ICRC07Dawson} for details).
%
\section{Discussion and outlook}
When inferring the energy spectrum from SD data we utilise the constant
intensity method to calibrate the SD data. The systematic uncertainties have
been scrutinised and the resulting spectrum is given. 
Several activities are on-going to reduce the systematic uncertainties of the
energy estimate, e.g.~the detector calibration uncertainty and the uncertainty
of the fluorescence yield. 
Reducing these uncertainties will make it desirable to 
deconvolve the energy spectrum 
using the estimate of the energy resolution.
The presented spectrum is compared with a spectrum derived
on basis of hybrid data only in T.~Yamamoto et
al.~\cite{bib:ICRC07Yamamoto}. Astrophysical implications are also discussed there.


\begin{thebibliography}{00}
\bibitem{bib:AugerNIM04} J.~Abraham [Pierre Auger Collaboration], NIM 523 (2004) 50. 
\bibitem{bib:RisseATP20} M.~Risse and D.~Heck, Astropart. Phys. 20 (2004) 661.
\bibitem{bib:Barbosa} H.~Barbosa et al., Astropart. Phys. 22 (2004) 159.
\bibitem{bib:ICRC07Dawson} B.~Dawson [Pierre Auger Collaboration] these
  proceedings, (2007), \#0976.
\bibitem{bib:ICRC07Suomijarvi} T.~Suomijarvi [Pierre Auger Collaboration] these
  proceedings, (2007) \#0299.  
\bibitem{bib:ICRC07Perrone} L.~Perrone [Pierre Auger Collaboration] these proceedings (2007), \#0316.
\bibitem{bib:ICRC05Barnhill} D.~Barnhill [Pierre Auger Collaboration],
  Proc. $29^{th}$ ICRC, Pune (2005), \textbf{7}, 291.
\bibitem{bib:ICRC05Allard} D.~Allard [Pierre Auger Collaboration],
  Proc. $29^{th}$ ICRC, Pune (2005), \textbf{7}, 71.
\bibitem{bib:Takeda03} M. Takeda et al., Astropart. Phys.~19 (2003) 447.
\bibitem{bib:ICRC05Sommers}P.~Sommers [Pierre Auger Collaboration],
  Proc. $29^{th}$ ICRC, Pune (2005) \textbf{7}, 387.
\bibitem{bib:ICRC07Ave}M.~Ave [Pierre Auger Collaboration] these
  proceedings, (2007), \#0297.
\bibitem{bib:ICRC07ben-zvi}S.~Ben-Zvi [Pierre Auger Collaboration] these
  proceedings, (2007) \#0399.
\bibitem{bib:JINST06Fick} B. Fick et al., JINST, 1 (2006) 11003.
\bibitem{bib:naganoFY} M. Nagano, K. Kobayakawa, N. Sakaki, K. Ando, Astropart. Phys. 22 (2004) 235.
\bibitem{bib:ICRC07Yamamoto}T.~Yamamoto [Pierre Auger Collaboration] these
  proceedings, (2007) \#0318. 
\end{thebibliography}
\end{document}